\documentclass[
reprint,
showpacs,preprintnumbers,
bibnotes,
 amsmath,amssymb,
 aps,
]{revtex4-1}

\usepackage{graphicx}
\usepackage{dcolumn}
\usepackage{bm}
\usepackage{subfigure}

\begin{document}

\title{Charged scalar perturbations around a  regular magnetic black hole }
\author{Yang Huang}
\email{saisehuang@163.com}
\author{Dao-Jun Liu}%
 \email{djliu@shnu.edu.cn}
 \affiliation{Center for Astrophysics, Shanghai Normal University, 100 Guilin Road, Shanghai 200234, China
}%
\date{\today}

\begin{abstract}
 We study charged scalar perturbations in the background of a   regular magnetic black hole.  In this case, the charged scalar perturbation does not result in superradiance. By using careful time-domain analysis, we show that the charge of scalar field can change the real part of quasinormal frequency, but has little impact on the imaginary part of  quasinormal frequency and the behavior of the late-time tail.  Therefore,  the regular magnetic black hole may be stable under the perturbations of charged scalar field at the linear level. 
   
\end{abstract}

\maketitle

\section{Introduction}

In classical general relativity, according to the Hawking-Penrose singularity theorem \cite{1970RSPSA.314..529H,Hawking:1973uf},  under the assumptions that the strong energy condition is valid and there exists  global hyperbolicity in the manifold of space-time, the collapse of matter inevitably produces a singularity. At a singularity, predictability is lost and
standard physics breaks down. Therefore, it is widely believed that space-time singularities do not exist in nature, but that they represent a limitation of the classical theory.  In the framework of black hole, the research of singularity-free black holes has attracted a considerable attention during the last few decades. The so-called regular black holes represent space-times
that have an event horizon and no pathological features like singularities or regions with closed timelike curves.  This kind of space-time is not a solution of Einstein's vacuum equations, but is introduced either with some exotic
field or modifications to gravity. We refer to \cite{Ansoldi:2008jw} for an interesting review on regular black holes; Ref.\cite{Lemos:2011dq}  contains also an review of the main results on regular black holes.     In particular, it is found that  when gravity couples to a suitable nonlinear
electrodynamics (NLED) field with the Lagrangian $L(F)$, ($F = -F_{\mu\nu} F^{\mu\nu}$), some regular black hole solutions with  electric charges \cite{AyonBeato:1998ub,AyónBeato199925,AyonBeato:1999ec} would be obtained \footnote{Actually,  a general static, spherically symmetric solution with an electric charge in GR coupled to NLED  has been obtained long before \cite{Pellicer1969}}. As a matter of fact, it has been proved in a general form that if the NLED has a Maxwell weak field limit, then the charged regular black hole solution in Ref. \cite{AyónBeato199925}  cannot have a true regular center \cite{BRONNIKOV197984}, unless the functional form of $L(F)$ is changed on some sphere \cite{Bronnikov2000}.  Bronnikov extended the solution to include radial magnetic field and showed that purely magnetic configurations can really have a regular center \cite{Bronnikov2001}.

Just like usual black hole solutions, the problem of stability for the regular black holes under linear perturbations is of interest. In fact, the scalar, Dirac, electromagnetic, and gravitational perturbations and the related quasinormal modes (QNMs) of the regular black hole have been  investigated in a variety of  models \cite{Flachi:2012nv,Li:2013fka,Toshmatov:2015wga,PhysRevD.86.064039,Xi:2016qrg}.  It is shown that the regular black holes are stable under various uncharged perturbations. 
As is well known, for the charged perturbations around a charged black hole,  the electromagnetic energy of the black hole can be superradiantly extracted and carried out by a wave with increased amplitude\cite{Bekenstein1973a}. It has been argued that charged Reissner-Nordström black holes and some charged black holes in string theory are stable to charged massive scalar perturbations\cite{Hod20131489,Hod2015prd,Li:2013jna}. 

The purpose of this work, is to investigate the behavior of charged scalar perturbations in the vicinity of a regular magnetic black hole.  The paper is organized as follows. In Sec.\ref{sec:mmfcsp}, we introduce the background space-time and electromagnetic potential for the regular magnetic black hole and the dynamical equation for charged scalar perturbations in the background.   Section \ref{sec:TDA} is devoted to a detailed numerical time-domain analysis on the evolution of the perturbations. Finally, we conclude our results in Sec. \ref{sec:conclution}. Throughout the paper, we use natural units with $c=\hbar=G=1$.

\section{METRIC, magnetic field and charged scalar PERTURBATION }
\label{sec:mmfcsp}
We shall consider a charged massive scalar field perturbation
in a static, spherically symmetric  regular magnetic black hole background with a  metric
\begin{equation}
	ds^2=-f(r)dt^2+\frac{1}{f(r)}dr^2+r^2d\theta^2+r^2\sin^2\theta d\phi^2,
\end{equation}
where the lapse function
\begin{equation}
	f(r)=1-\frac{2M}{r}\left[1-\tanh\left(\frac{q^2_m}{2Mr}\right)\right].
\end{equation}
Here, the integration constants $M$ and $q_m$ denote the  mass and magnetic charge of the black hole, respectively.  As was shown in Ref.\cite{Bronnikov2001}, the electromagnetic field $F_{\mu\nu}=\partial_{\mu}A_{\nu}-\partial_{\nu}A_{\mu}$ in the background of a regular magnetic black hole only involves a radial magnetic field
\begin{equation}
	F_{23}=-F_{32}=q_m\sin\theta.
\end{equation}
One can easily show that the electromagnetic potential takes the form
\begin{equation}
	A_\mu=(0,0,0,-q_m\sin\theta).
\end{equation}

The equation of motion for a charged scalar field with mass $\mu$ and electric charge $q$ in the background of a regular magnetic black hole described above reads
\begin{equation}\label{Klein Gordon Eq-1}
	\left(\nabla_\mu-i q A_\mu\right)\left(\nabla^\mu-i q A^\mu\right)\Psi=\mu^2\Psi.
\end{equation}
The above equation can be explicitly written as
\begin{equation}\label{Klein Gordon Eq-2}
	\begin{aligned}
		&-\frac{1}{f(r)}\frac{\partial^2\Psi}{\partial t^2}+\frac{1}{r^2}\frac{\partial}{\partial r}\left(r^2f(r)\frac{\partial}{\partial r}\Psi\right)+\frac{1}{r^2\sin\theta}\frac{\partial}{\partial\theta}\left(\sin\theta\frac{\partial}{\partial\theta}\Psi\right)\\&+\frac{1}{r^2\sin^2\theta}\left(\frac{\partial}{\partial\phi}+i\alpha\cos\theta\right)^2\Psi=\mu^2\Psi
	\end{aligned}
\end{equation}
where $\alpha\equiv q q_m$, which describes the electromagnetic interaction between the charged scalar perturbation and the regular black hole. For the case of  $q_m=0$ , Eq.(\ref{Klein Gordon Eq-2}) is just the equation of motion for  the massive scalar field  in the background of Schwarzschild space-time.  When $q$ is set to be zero, Eq.(\ref{Klein Gordon Eq-2}) is reduced to the equation that describes a neutral scalar field in a regular black hole background.   Contrary to the case of the electrically charged regular black hole, the  equation for a charged scalar perturbation around regular magnetic black hole seems to be inseparable, due to the term containing $\alpha$.  However, one can expand the field equation in a complete basis of spherical harmonics and obtain a set of 1+1 dimensional equations with $\ell$-mode-mixing couplings by making use of the orthogonality properties of the spherical harmonics. This method has been deeply discussed in the literature \cite{PhysRevD.86.104017,2013-Dolan-p124026-124026,Pani:2013pma}. 

We decompose the wave equation (\ref{Klein Gordon Eq-1}) with the ansatz
\begin{equation}\label{ansatz}
	\Psi(t,r,\theta,\phi)=\sum_{\ell m}Y_{\ell m}(\theta,\phi)\frac{\psi_{\ell m}(t,r)}{r}
\end{equation}
where $Y_{\ell m}(\theta,\phi)$ denotes spherical harmonics. Following \cite{Pani:2013pma}, we will append the relevant multipolar index $\ell$ to any perturbation variable but we will omit the index $m$, because in an spherically symmetric background it is possible to decouple the perturbation equations so that all quantities have the same value of $m$. Thus we have $Y^{\ell}\equiv Y_{\ell m}(\theta,\phi),\;\psi_{\ell}(t,r)\equiv \psi_{\ell m}(t,r)$. Using the ansatz (\ref{ansatz}) , we obtain the following equation from  Eq.(\ref{Klein Gordon Eq-2})
\begin{equation}\label{eq:general equation}
	A_\ell Y^{\ell}+2m\alpha\varpi(r)\psi_\ell\cos\theta Y^{\ell}+\left[\alpha^2\varpi(r)\psi_\ell-A_\ell\right]\cos^2\theta Y^{\ell}=0,
\end{equation}
where
\begin{equation}
	\begin{aligned}
		&A_\ell=\left[\partial_{tt}-\partial_{xx}+V_{\ell}\right]\psi_\ell(t,r),\\
		&\varpi(r)=\frac{f(r)}{r^2},
	\end{aligned}
\end{equation}
and the potential 
\begin{equation}\label{Aell Bell & Cell}
		V_{\ell}=f(r)\left[\frac{f'(r)}{r}+\frac{\ell(\ell+1)}{r^2}+\mu^2\right].
\end{equation}
Here $x$ is the tortoise coordinate defined by $dx=dr/f(r)$.  Note that a sum over $m$ and $\ell \geq |m|$ in Eq.(\ref{eq:general equation}) is implicit. For the neutral perturbations, that is, $\alpha=0$,  Eq.(\ref{eq:general equation}) is reduced to  a single second-order differential equation for a given value of $\ell$
\begin{equation}
	A_{\ell}=\left[\partial_{tt}-\partial_{xx}+V_{\ell}\right]\psi_\ell(t,r)=0.
\end{equation} 
For the charged perturbations, however, the $\cos\theta$ and $\cos^2\theta$ terms in Eq.(\ref{eq:general equation}) will lead inevitably to some coupling between different $\ell$ modes. By  using  the  orthogonality properties of scalar spherical harmonics
\begin{equation}\label{the orthogonality property}
	\int Y^\ell Y^{*\ell'}d\Omega=\delta^{\ell\ell'}
\end{equation}
and the identities \cite{Kojima1992,Pani:2013pma}
\begin{equation}\label{identities-1}
	\begin{aligned}
		&\cos\theta Y^{\ell}=\mathcal{Q}_{\ell+1}Y^{\ell+1}+\mathcal{Q}_{\ell}Y^{\ell-1},
		\\\cos^2\theta Y^{\ell}=
		&\left(\mathcal{Q}^2_{\ell+1}+\mathcal{Q}^2_{\ell}\right)Y^{\ell}+\mathcal{Q}_{\ell+1}\mathcal{Q}_{\ell+2}Y^{\ell+2}\\&+\mathcal{Q}_{\ell}\mathcal{Q}_{\ell-1}Y^{\ell-2},
	\end{aligned}
\end{equation}
where $Q_\ell$ is defined as
\begin{equation}
	\mathcal{Q}_\ell=\sqrt{\frac{\ell^2-m^2}{4\ell^2-1}},
\end{equation}
and integrating  Eq.(\ref{eq:general equation}) over the two-sphere, we obtain the following equation:
\begin{widetext}
	\begin{equation}\label{coupled eqs}
		\begin{aligned}
		-\mathcal{Q}_{\ell-1}\mathcal{Q}_{\ell}A_{\ell-2}&+\left(1-\mathcal{Q}^2_{\ell+1}-\mathcal{Q}^2_{\ell}\right)A_{\ell}-\mathcal{Q}_{\ell+1}\mathcal{Q}_{\ell+2}A_{\ell+2}=-2m\alpha\varpi(r)\left(\mathcal{Q_\ell}\psi_{\ell-1}+\mathcal{Q}_{\ell+1}\psi_{\ell+1}\right)\\&-\alpha^2\varpi(r)\left[\mathcal{Q}_{\ell-1}\mathcal{Q}_{\ell}\psi_{\ell-2}+\left(\mathcal{Q}^2_{\ell+1}+\mathcal{Q}^2_{\ell}\right)\psi_{\ell}+\mathcal{Q}_{\ell+1}\mathcal{Q}_{\ell+2}\psi_{\ell+2}\right].
		\end{aligned}
	\end{equation}
\end{widetext}
Clearly, the angular dependence has been completely eliminated and the problem has been reduce to a $(1+1)$-dimensional equation. 
For the case $m=0$, the even-$\ell$ and odd-$\ell$ sectors are completely decoupled and belong to two separated sets of equations. In this case, it is noteworthy that perturbations with harmonic index $\ell$ are only coupled with perturbations with $\ell\pm2$. On the other hand, when $m\neq0$, perturbations with  index $\ell$ are not only  coupled with perturbations with $\ell\pm2$, but also with those labeled by $\ell\pm1$. It should be pointed out that Eq.(\ref{coupled eqs}) actually contains an infinite number of coupled equations, which is similar in some regards to that developed in Ref.\cite{Csizmadia2013On,Pani:2013pma}.  In practical calculations, the above coupled system can be solved by truncating the sum over harmonic indices $\ell$ to some order $L$, and checking convergence of the results when $L$ is sufficiently large \cite{2013-Dolan-p124026-124026}.  

\section{Time-domain analysis}
\label{sec:TDA}

Time-domain analysis offer an  intuitive approach to study the stability and/or instability of black hole under linear perturbations. The criterion to determine whether a black hole is stable or not is whether the time-domain profile for the evolution of the perturbation is decaying or not. Time-domain profiles include contributions from all possible modes.

\subsection{Method}

 To be precise, let us rewrite Eq.(\ref{coupled eqs}) in a simpler form
\begin{equation}\label{coupled eqs matrix form}
	\begin{aligned}
	\left(\partial_{tt}-\partial_{xx}\right)\vec{\psi}+\mathbf{V}(r)\vec{\psi}=0
	\end{aligned}
\end{equation}
where 
\begin{equation}\label{potential matrix}
	\mathbf{V}(r)=\mathrm{diag}[{\vec{V}}]+2m\alpha\varpi(r)\mathbf{A}^{-1}\mathbf{B}+\alpha^2\varpi(r)\left(\mathbf{A}^{-1}-\mathbf{I}\right).
\end{equation}
Here, $\mathbf{A}$ and $\mathbf{B}$ are two constant  matrices,  $\mathbf{I}$ is an identity matrix and $\mathrm{diag}[{\vec{V}}]$ represents a diagonal matrix  having $\vec{V}$ as its main diagonal. Vectors $\vec{V}$ and $\vec{\psi}$ in Eq.(\ref{coupled eqs matrix form}) are defined as
\begin{equation}
	\vec{V}=(V_{k},V_{k+1},V_{k+2},\cdots)^{T},\quad\vec{\psi}=(\psi_{k},\psi_{k+1},\psi_{k+2},\cdots)^{T},
\end{equation}
respectively, where $V_{k}$'s are the potentials given in Eq.(\ref{Aell Bell & Cell}) and the subscript $k$ 
\begin{equation}\label{eqn:k}
	k=\left\{\
	\begin{aligned}
	&0\;\mathrm{or}\;1,\quad m=0,\\&|m|,\quad m\neq0.
	\end{aligned}
	\right.
\end{equation} 
The explicit form of matrices $\mathbf{A}$ and $\mathbf{B}$ in Eq.(\ref{potential matrix}) reads
\begin{widetext}
	\begin{equation}
	\mathbf{A}=\left(
	\begin{array}{cccccc}
	1-\mathcal{Q}^2_k-\mathcal{Q}^2_{k+1} & 0 & -\mathcal{Q}_{k+1}\mathcal{Q}_{k+2} & 0 & 0 & \cdots \\
	0 & 1-\mathcal{Q}^2_{k+1}-\mathcal{Q}^2_{k+2} & 0 & -\mathcal{Q}_{k+2}\mathcal{Q}_{k+3} & 0 & \cdots \\
	-\mathcal{Q}_{k+1}\mathcal{Q}_{k+2} & 0 & 1-\mathcal{Q}^2_{k+2}-\mathcal{Q}^2_{k+3} & 0 & -\mathcal{Q}_{k+3}\mathcal{Q}_{k+4} & \cdots \\
	0 & -\mathcal{Q}_{k+2}\mathcal{Q}_{k+3} & 0 & 1-\mathcal{Q}^2_{k+3}-\mathcal{Q}^2_{k+4} & 0 & \cdots \\
	0 & 0 & -\mathcal{Q}_{k+3}\mathcal{Q}_{k+4} & 0 & 1-\mathcal{Q}^2_{k+4}-\mathcal{Q}^2_{k+5} & \cdots \\
	\cdots & \cdots & \cdots & \cdots & \cdots & \cdots \\
	\end{array}
	\right),
	\end{equation}
	\begin{equation}
	\mathbf{B}=\left(
	\begin{array}{cccccc}
	0 & \mathcal{Q}_{k+1} & 0 & 0 & 0 & \cdots \\
	\mathcal{Q}_{k+1} & 0 & \mathcal{Q}_{k+2} & 0 & 0 & \cdots \\
	0 & \mathcal{Q}_{k+2} & 0 & \mathcal{Q}_{k+3} & 0 & \cdots \\
	0 & 0 & \mathcal{Q}_{k+3} & 0 & \mathcal{Q}_{k+4} & \cdots \\
	0 & 0 & 0 & \mathcal{Q}_{k+4} & 0 & \cdots \\
	\cdots & \cdots & \cdots & \cdots & \cdots & \cdots \\
	\end{array}
	\right).
	\end{equation}
\end{widetext}

We will extend the technique developed by Gundlach, Price and Pullin \cite{Gundlach1994prd} to integrate Eq.(\ref{coupled eqs matrix form}). To this end, we first use the light-cone coordinates $du=dt-dx$ and $dv=dt+dx$,  and rewrite Eq.(\ref{coupled eqs matrix form}) as
\begin{equation}\label{eqsuv}
	\begin{aligned}
	4\frac{\partial^2}{\partial u\partial v}\vec{\psi}+\mathbf{V}(r)\vec{\psi}=0.
	\end{aligned}
\end{equation}

Then we discretize the vectorized wave equation (\ref{coupled eqs matrix form})
 and obtain that
\begin{equation}
	\begin{aligned}
	\vec{\psi}(N)=&\vec{\psi}(W)+\vec{\psi}(E)-\vec{\psi}(S)\\&-\frac{h^2}{8}\mathbf{V}(S)\left[\vec{\psi}(W)+\vec{\psi}(E)\right]+\mathcal{O}(h^4),
	\end{aligned}
\end{equation}
where $N$, $W$, $E$ and $S$ are the points of a unit grid on the $u-v$ plane which correspond  to the points $(u+h, v+h)$, $(u+h,v)$, $(u, v+h)$ and $(u,v)$, respectively. Here $h$ is step length of the change of 
$u$ or $v$. 

Besides, we assume the field $\psi$ initially takes  the form as a Gaussian wave package
\begin{equation}\label{eqn:GWP}
\psi_{\ell}(v, u=0)=\exp\left[\frac{-(v-v_{\ast})^2}{2\sigma^2}\right]
\end{equation}
for a particular $\ell$-mode.

After numerically integrating Eq.(\ref{eqsuv}), we obtain the time-domian profile, which is a series of values of the perturbation field $\vec{\psi}(t=(v+u)/2,x=(v-u)/2)$  at a given position $x$ (we always set $x=10M$ in our computations) and discrete  moments $t=t_0=(v_0+u_0)/2,t=t_0+h, t=t_0+2h,\cdots$.

What we are interested in  is the behavior of perturbations in two stages. The first one is a period  of damping proper oscillations which follows a relatively short period of initial outburst of radiation and is dominated by QNMs. 
From the time-domain profile, we can extract dominant frequency of QNMs  with the help of the Prony method (see, e.g., Ref. \cite{berti_mining_2007}). That is, we can fit the profile data by a superposition of damping or growing exponents
 \begin{equation}
 	\label{Eq:Prony1}
 	\psi(t, x)\simeq \sum_{i=1}^{p}C_ie^{-i \omega_i(t-t_0)}.
 \end{equation}
We start at a relative late moment $t_0$ and end at $t=t_0+N h$, where $N$ is an integer and $N \geq 2p-1$. It is of importance to ensure that both $t_0$ and $t$ are located in the QNM-dominant stage.  Then the formula (\ref{Eq:Prony1}) is valid for each value from the profile data:
\begin{equation}
y_n\equiv\psi(t_0+n h,x)=\sum_{j=1}^{p}C_je^{-i\omega_j n h}=\sum_{j=1}^{p}C_{j}z_j^n.
\end{equation}
The Prony method allows us to find $z_j$ in terms of the known $y_n$ and to calculate the frequencies $\omega_j$ because $h$ has been given. In fact, because these frequencies are usually complex numbers, it is  the real part of  $\omega_j$ that denotes the oscillating frequency, while the imaginary part denotes damping rate if it is negative and growth rate if positive.    The dominant mode is selected as the one that has the greatest amplitude during the period  among the $p$  modes we consider. 

Another stage we are also interested in is the late-time tail period. 
From the time-domain profile, we can  fit the tail by directly using the following formula
\begin{equation}\label{tailfitting}
	\psi(t,x)=c_T\exp(\lambda_T t)\cos(\omega_T t+\phi_T),
\end{equation}  
where $c_T, \lambda_T, \omega_T$ and $\phi_T$ are four free parameters. What interests us are the values of $\lambda_T$ and $\omega_T$, which describe the slope and the oscillation frequency of the tail in the $\ln|\psi|-t$ plane. 

Without loss of generality,  we always choose the  magnetic charge of the black hole  $q_m=0.5M$  and choose the parameters $v_{\ast}=1$ and $\sigma=1$ for the initial Gaussian wave package in the practical computation. 

Before we present our numerical results, we first check that the results we obtained are robust under adjustments of the maximum number of $\ell$ modes used. Fortunately, the couplings die away rapidly with increasing $\ell$ and we find that for the values of $q M\lesssim 1$, it is enough to consider only three to five coupled $\ell$ modes. Therefore, the method is useful in practice. In what following, we always take five coupled $\ell$-modes in the computation. 

\subsection{Numerical Results}

\begin{figure*}    
	\subfigure { \label{fig:e}     
		\includegraphics[width=1\columnwidth]{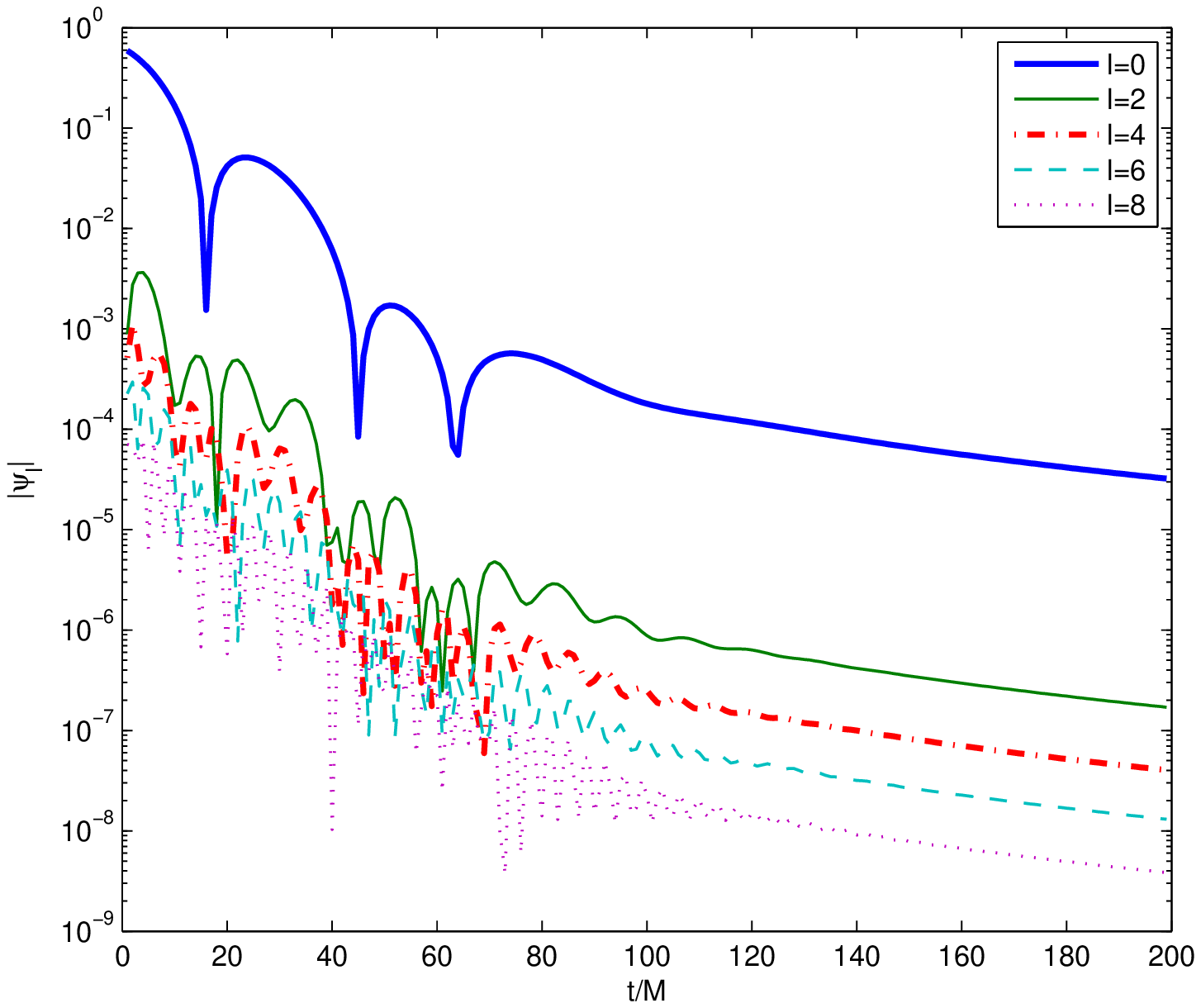} 
	}     
	\subfigure { \label{fig:f}     
		\includegraphics[width=1\columnwidth]{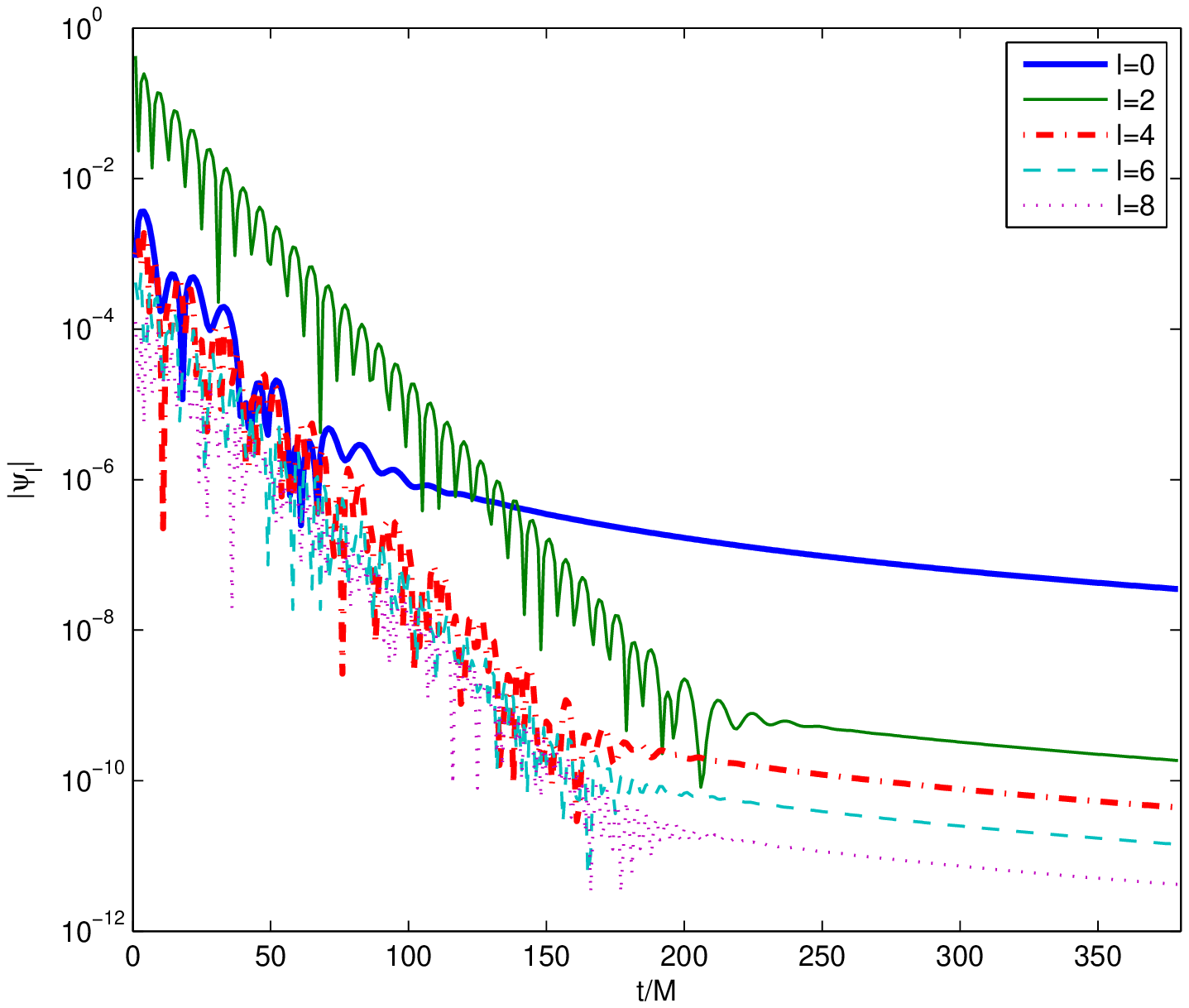}  }
	\caption{The time-domain profiles of five different  $\ell$ modes with  magnetic quantum number $m=0$ for a massless perturbation. Here we take the charge of the scalar perturbation $q M=0.2$. An initial Gaussian wave package described in Eq.(\ref{eqn:GWP})  is set to the $\ell=0$ mode and $\psi_{\ell\neq0}(v,u=0)=0$ for the left panel; for the right panel, the same initial Gaussian wave package is set to the $\ell=2$ mode and $\psi_{\ell\neq2}(v,u=0)=0$.}
	\label{fig:1}     
\end{figure*}

First, we present the time-domain profiles of different  $\ell$ modes  with  magnetic quantum number $m=0$ for a massless perturbation in Fig.\ref{fig:1} of which the charge  is $q M=0.2$. For comparison, we take the initial Gaussian wave package described in Eq.(\ref{eqn:GWP}) to two different $\ell$ modes in the left panel and the right, respectively.
In the left panel,  the ($\ell=0$) mode has the initial waveform of Gaussian wave package and $\psi_{\ell\neq0}(v,u=0)=0$. It is obvious that the amplitude of the ($\ell=0$) mode is much greater than those of other $\ell$-modes and the  greater the multipolar index $\ell$ is, the less value the amplitude takes. This is not difficult to understand, because all the ($\ell\neq 0$)-modes are initially absent and their amplitude are induced one by one from the  ($\ell=0$)-mode  due to the coupling term containing parameter $\alpha$ which is small enough to make the inducing effect weaken rapidly. 
However, in the right panel, the same initial Gaussian wave package is given to ($\ell=2$)-mode and $\psi_{\ell\neq2}(v,u=0)=0$.  As we expect, the ($\ell=2$)-mode initially dominates the profiles and the amplitudes of the  $\ell=0$, $\ell=4$, $\ell=6$ and $\ell=8$ modes are  reduced one by one.  It is interesting to find that the behavior of ($\ell=0$)-mode in the right panel is just the same as that of ($\ell=2$)-mode in the left panel. This reflects the fact that the inducing effect between   two adjacent $\ell$-modes is mutual.  In the late-time period, the tail of  ($\ell=0$)-mode   exceeds that of ($\ell=2$)-mode and dominates the profiles.  Furthermore, the tails of all $\ell$-modes take the same slope, which  may be explained by considering that after a long period of mutual interaction, all the modes reach  equilibrium.   From this we may infer that the slope of the tails has little relevance to the multipole index $\ell$.    Besides, we want to point out that,  relative to  the initially-existing mode,  in  the QNM-dominating period, the purely induced $\ell$-modes are heavily impacted by their adjacent  modes.  Therefore,  we will obtain  more accurate results for the initially-existing mode when we  use Prony method to compute the frequencies of QNMs. In what follows, we will only present the profiles of the ($\ell=k$)-mode, where $k$ is determined by Eq.(\ref{eqn:k})， and in the detailed  computation, we will always set an initial Gaussian wave package to the ($\ell=k$) mode. 

\begin{figure*}    
 	\subfigure { \label{fig:a}     
 		\includegraphics[width=\columnwidth]{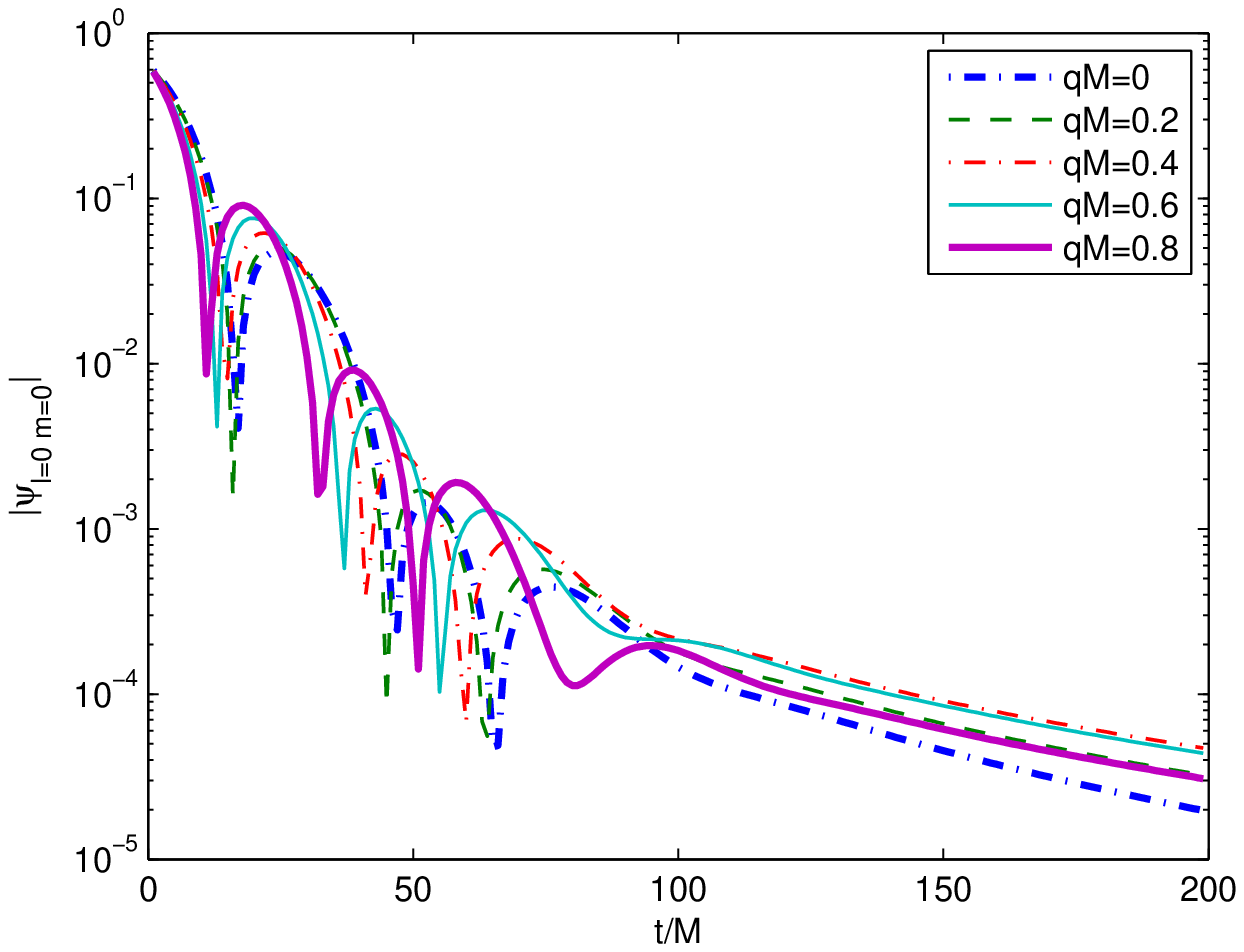} 
 	}     
 	\subfigure { \label{fig:b}     
 		\includegraphics[width=\columnwidth]{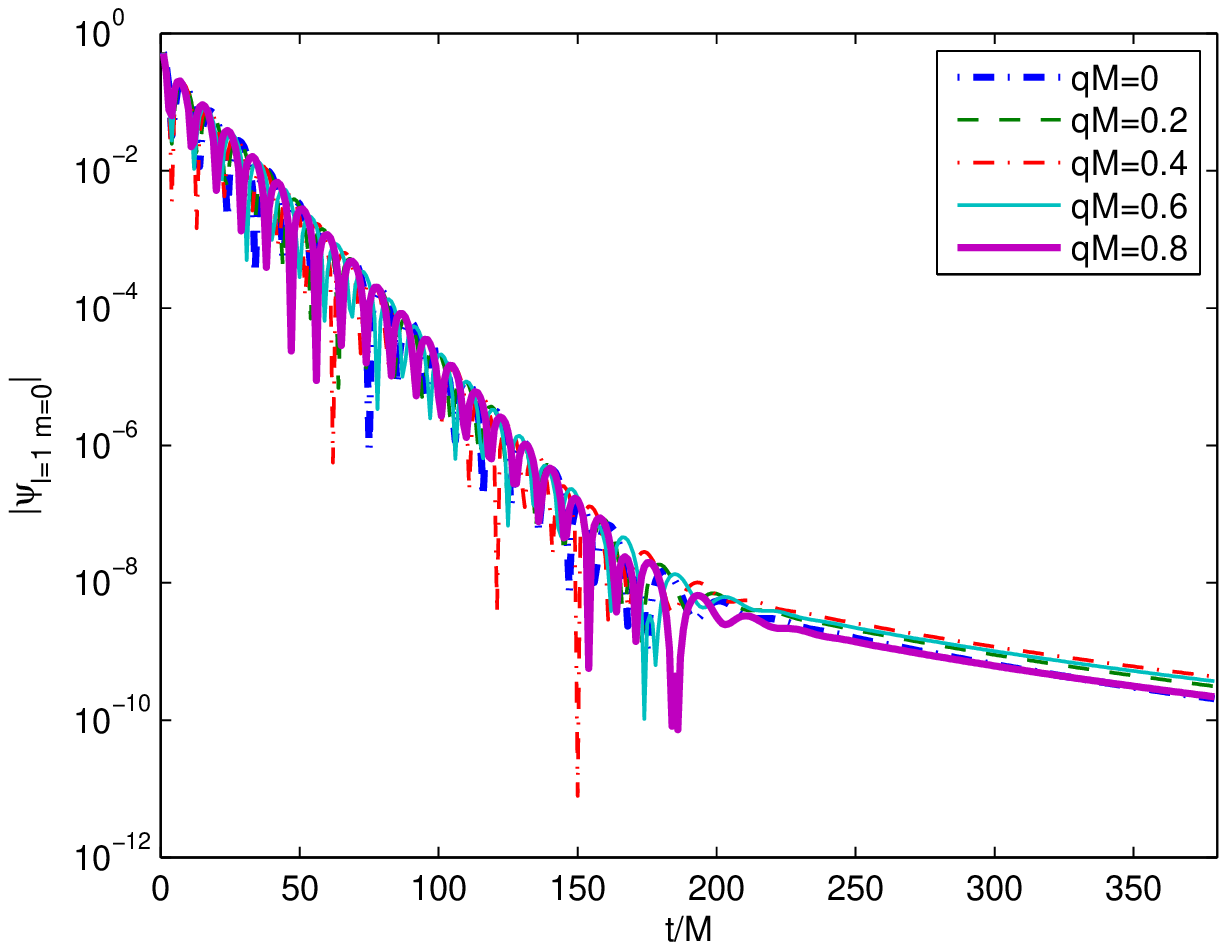}  }
 	\caption{
 		The time-domain profiles with  magnetic quantum number $m=0$ for  massless perturbations with different charge. Here we plot the profiles for the ($\ell=0$) mode and ($\ell=1$) mode in the left and right panel, respectively.
 	}
 	\label{fig:5_6}     
 \end{figure*}

 Figure {\ref{fig:5_6}} shows  the time-domain profiles with  magnetic quantum number $m=0$ for  massless perturbations with different charge. The profiles for ($\ell=0$)-mode and ($\ell=1$)-mode are plotted in the left and right panel, respectively. The frequencies of dominant QNM corresponding to the profiles are given in Table \ref{table:1}. The real part of the frequencies for the ($\ell=0$)-mode and ($\ell=1$)-mode is found with the increase of perturbation's charge $q$. However, although the imaginary part is always negative indicating the stability of the perturbation, its absolute value becomes slightly smaller.    
 \begin{table}
	\caption{The frequency of dominant QNMs corresponding to each profile presented in Fig.\ref{fig:5_6}.}\label{table:1}
	\begin{tabular}{ccc}\hline\hline
		\centering
		$qM$\;\;\;&$m=0\;\ell=0$\;\;\;& $m=0\;\ell=1$\\\hline
		0 \;\;\; & $ 0.115986-0.106589i $\;\;\;&$0.307023-0.0983725i$\\
		0.2\;\;\; &$ 0.127791-0.105843i $\;\;\;&$0.310334 -0.0983376i$\\
		0.4\;\;\; &$ 0.128022-0.104195i $\;\;\;&$0.319776 -0.0982426i$\\
		0.6\;\;\; &$ 0.139425-0.103115i $\;\;\;& $0.334088-0.0981062i $ \\
		0.8\;\;\; &$ 0.1543-0.102291i $\;\;\;& $ 0.35161-0.0979471i $ \\
		\hline\hline
	\end{tabular}
\end{table}

\begin{figure*}    
	\subfigure{\includegraphics[width=\columnwidth]{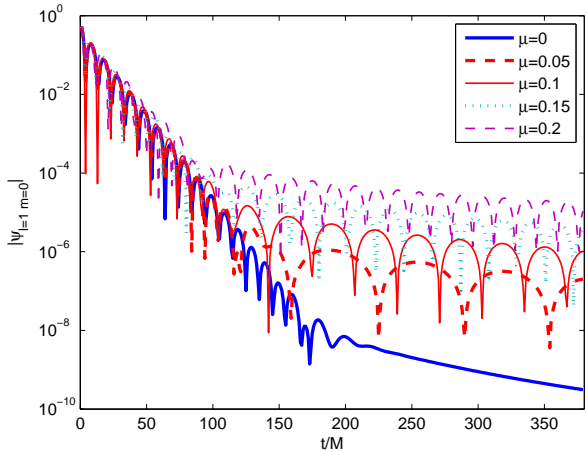}}
	\subfigure { \label{fig:g}     
		\includegraphics[width=\columnwidth]{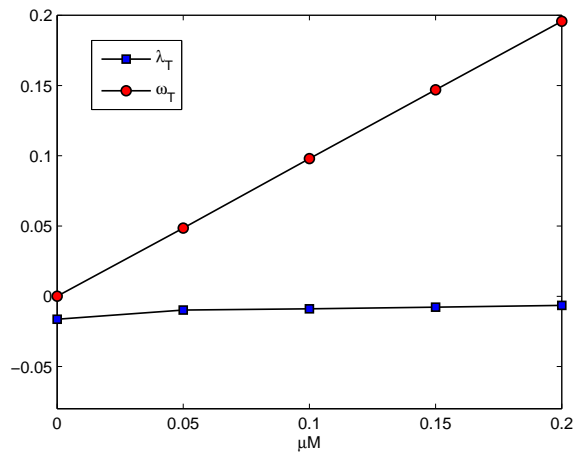} 
	}
	\caption{
		Left: The time-domain profiles of ($\ell=1$)-mode with  magnetic quantum number $m=0$ for 5 scalar perturbations with the same charge ($q M= 0.2$) and different mass $\mu$; Right: The slope ($\lambda_T$) and the oscillation frequency ($\omega_T$) of the late-time tails for profiles plotted in the left panel.
		}
	\label{fig:44}     
\end{figure*}

In the left panel of Fig.\ref{fig:44}, we plot the time-domain profiles of ($\ell=1$)-mode with  magnetic quantum number $m=0$ for scalar perturbations with the same charge $q$ and different mass $\mu$.
The mass has little impact on the slope of the late-time tail, at least for the values of mass $\mu$ that we consider. However, the mass of scalar field perturbation can make the tail oscillate and the frequency is accurately proportional to the mass value, as is shown in the right panel of Fig. \ref{fig:44}. The mass of the scalar perturbation will change distinctly the value of quasinormal frequency. From table \ref{table:3}, we can see that the real part of the frequency is found with the increase of the mass value, while the absolute value of the  imaginary part becomes smaller. Obviously but importantly, the mass does not change the damping nature of the QNMs. 
\begin{table}
	\caption{The frequency of dominant QNMs corresponding to each profile presented in the left panel of Fig.\ref{fig:44}.}\label{table:3}
	\begin{tabular}{cc}\hline\hline
		\centering
		$\mu M$\;\;\;&$\omega M$\\\hline
		0 \;\;\; & $0.309805-0.0985177i$\\
		0.05\;\;\; &$0.310986-0.0978657i$\\
		0.1\;\;\; &$0.314368-0.0969102i$\\
		0.15\;\;\; &$0.321369-0.0919812i$\\
		0.2\;\;\; &$0.329841-0.0896732i$\\
		\hline\hline
	\end{tabular}
\end{table}

\begin{figure}    
	\includegraphics[width=\columnwidth]{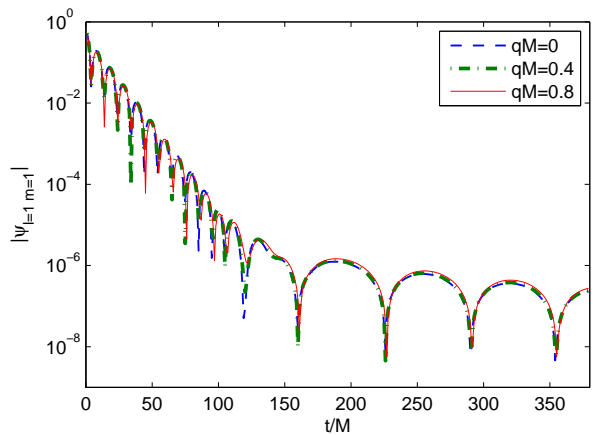}
	\caption{
		The time-domain profiles of ($\ell=1$)-mode with  magnetic quantum number $m=1$ for some scalar perturbations with the same mass ($\mu M= 0.05$) and different charge $q$.
        }
	\label{fig:000}     
\end{figure}

The time-domain profiles of ($\ell=1$)-mode with  magnetic quantum number $m=1$ for some scalar perturbations with the same mass ($\mu M= 0.05$) and different charge $q$ are plotted in Fig.\ref{fig:000}.
When  magnetic quantum number $m=1$,  the ($\ell=1$)-mode interacts not only with the ($\ell=3$)-mode, but also with the ($\ell=2$)-mode. Therefore, the impact of the perturbation's  charge on the QNM and the late-time tail may be different from the one in the case of $m=0$ that we have described. In fact, from Fig.\ref{fig:000} and Table \ref{table:2}, we can find that,  except that the real part of frequency of QNMs become slightly small with the increase of the charge $q$,  the late-time tail and the imaginary part of QNMs' frequency are hardly influenced by the charge. 
\begin{table}
	\caption{The dominant quasinormal frequency $\omega$ as well as the tail's slope and oscillation frequency  of the ($\ell=1$)-mode with  magnetic quantum number $m=1$ for a scalar perturbation with the same mass ($\mu M= 0.05$) and different charge $q$.}\label{table:2}
	\begin{tabular}{cccc}\hline\hline
		\centering
		$qM$\;\;&$\omega M$\;\;&$\omega_T M$\;\;\;& $\lambda_T M$\\\hline
		0 \;\;  &$ 0.308064 -0.0977589i $\;\;&$0.048686$\;\;&$-0.0104283$\\
		0.2\;\; &$ 0.307890 -0.0979463i $\;\;&$0.048677$\;\;&$-0.0104251$\\
		0.4\;\; &$ 0.306895 -0.0977342i $\;\;&$0.048698$\;\;&$-0.0104098$\\
		0.6\;\; &$ 0.305129 -0.0976514i $\;\;&$0.048721$\;\;&$-0.0103680$\\
		0.8\;\; &$ 0.303128 -0.0976198i $\;\;&$0.048753$\;\;&$-0.0103232$\\
		\hline\hline
	\end{tabular}
\end{table}

\section{Conclusion}
\label{sec:conclution}

In this paper, the behavior of time evolution for charged scalar perturbations in the background of a   regular magnetic black hole was investigated. It was found that, radically different from the charged perturbations around the electrically charged regular black hole with the same outside metric, there exist some  interactions between the different $\ell$ modes. This kind of interaction is  related to the magnetic charge of the black hole and the electric charge of the perturbations. Moreover,  the charged scalar perturbation here shows no  superradiance  behavior. This can be understood from Eq.(\ref{coupled eqs matrix form}), where the matrix-valued potential $\mathbf{V}(r)$ vanishes gradually both at the neighborhood of the event horizon of the black hole and at infinity.  In fact, the absence of superradiance can also be seen directly from Eq.
(\ref{Klein Gordon Eq-2}), since it does not meet the general requirements for superradiance described in Ref.\cite{Richartz:2009mi}. Notice that this is crucially related to the magnetic character of the black hole charge. Had we considering an electrically charged black hole, superradiance would be present.

By using careful time-domain analysis, we show that the charge of scalar field can change the real part of quasinormal frequency, but has little impact on the imaginary part of  quasinormal frequency and the behavior of the late-time tail, at least for the values  of the charge we take into consideration.  Therefore,  the regular magnetic black hole may be stable under the perturbations of charged scalar field at the linear level. 
Physically, there would be a simple explanation for this stability: since the electromagnetic field of the regular black hole has only a magnetic component, it will not do any work on the test field with or without charge and, then,  it is reasonable to 
expect that the results about stability will not be very different from those for uncharged fields.

It is of interest to calculate the quasinormal frequency  up to a high value of overtone number. To this end , a detailed frequency-domain analysis will be helpful, which deserves new work in the furture.

\begin{acknowledgments}
We are grateful to Prof. K. A. Bronnikov and Ran Li for helpful correspondence.
This work is supported in part by National Natural Science Foundation of China under Grant  No.~11275128, Shanghai Municipal Commission of Science and technology under Grant No.~12ZR1421700 and the Program of Shanghai Normal University.
\end{acknowledgments}

\bibliography{myRef}

\begin{thebibliography}{29}%
\makeatletter
\providecommand \@ifxundefined [1]{%
 \@ifx{#1\undefined}
}%
\providecommand \@ifnum [1]{%
 \ifnum #1\expandafter \@firstoftwo
 \else \expandafter \@secondoftwo
 \fi
}%
\providecommand \@ifx [1]{%
 \ifx #1\expandafter \@firstoftwo
 \else \expandafter \@secondoftwo
 \fi
}%
\providecommand \natexlab [1]{#1}%
\providecommand \enquote  [1]{``#1''}%
\providecommand \bibnamefont  [1]{#1}%
\providecommand \bibfnamefont [1]{#1}%
\providecommand \citenamefont [1]{#1}%
\providecommand \href@noop [0]{\@secondoftwo}%
\providecommand \href [0]{\begingroup \@sanitize@url \@href}%
\providecommand \@href[1]{\@@startlink{#1}\@@href}%
\providecommand \@@href[1]{\endgroup#1\@@endlink}%
\providecommand \@sanitize@url [0]{\catcode `\\12\catcode `\$12\catcode
  `\&12\catcode `\#12\catcode `\^12\catcode `\_12\catcode `\%12\relax}%
\providecommand \@@startlink[1]{}%
\providecommand \@@endlink[0]{}%
\providecommand \url  [0]{\begingroup\@sanitize@url \@url }%
\providecommand \@url [1]{\endgroup\@href {#1}{\urlprefix }}%
\providecommand \urlprefix  [0]{URL }%
\providecommand \Eprint [0]{\href }%
\providecommand \doibase [0]{http://dx.doi.org/}%
\providecommand \selectlanguage [0]{\@gobble}%
\providecommand \bibinfo  [0]{\@secondoftwo}%
\providecommand \bibfield  [0]{\@secondoftwo}%
\providecommand \translation [1]{[#1]}%
\providecommand \BibitemOpen [0]{}%
\providecommand \bibitemStop [0]{}%
\providecommand \bibitemNoStop [0]{.\EOS\space}%
\providecommand \EOS [0]{\spacefactor3000\relax}%
\providecommand \BibitemShut  [1]{\csname bibitem#1\endcsname}%
\let\auto@bib@innerbib\@empty
\bibitem [{\citenamefont {{Hawking}}\ and\ \citenamefont
  {{Penrose}}(1970)}]{1970RSPSA.314..529H}%
  \BibitemOpen
  \bibfield  {author} {\bibinfo {author} {\bibfnamefont {S.~W.}\ \bibnamefont
  {{Hawking}}}\ and\ \bibinfo {author} {\bibfnamefont {R.}~\bibnamefont
  {{Penrose}}},\ }\href {\doibase 10.1098/rspa.1970.0021} {\bibfield  {journal}
  {\bibinfo  {journal} {Royal Society of London Proceedings Series A}\ }\textbf
  {\bibinfo {volume} {314}},\ \bibinfo {pages} {529} (\bibinfo {year}
  {1970})}\BibitemShut {NoStop}%
\bibitem [{\citenamefont {Hawking}\ and\ \citenamefont
  {Ellis}(2011)}]{Hawking:1973uf}%
  \BibitemOpen
  \bibfield  {author} {\bibinfo {author} {\bibfnamefont {S.~W.}\ \bibnamefont
  {Hawking}}\ and\ \bibinfo {author} {\bibfnamefont {G.~F.~R.}\ \bibnamefont
  {Ellis}},\ }\href@noop {} {\emph {\bibinfo {title} {{The Large Scale
  Structure of Space-Time}}}},\ Cambridge Monographs on Mathematical Physics\
  (\bibinfo  {publisher} {Cambridge University Press},\ \bibinfo {year}
  {2011})\BibitemShut {NoStop}%
\bibitem [{\citenamefont {Ansoldi}(2008)}]{Ansoldi:2008jw}%
  \BibitemOpen
  \bibfield  {author} {\bibinfo {author} {\bibfnamefont {S.}~\bibnamefont
  {Ansoldi}}\ }(\bibinfo {year} {2008})\ \Eprint
  {http://arxiv.org/abs/0802.0330} {arXiv:0802.0330 [gr-qc]} \BibitemShut
  {NoStop}%
\bibitem [{\citenamefont {Lemos}\ and\ \citenamefont
  {Zanchin}(2011)}]{Lemos:2011dq}%
  \BibitemOpen
  \bibfield  {author} {\bibinfo {author} {\bibfnamefont {J.~P.~S.}\
  \bibnamefont {Lemos}}\ and\ \bibinfo {author} {\bibfnamefont {V.~T.}\
  \bibnamefont {Zanchin}},\ }\href {\doibase 10.1103/PhysRevD.83.124005}
  {\bibfield  {journal} {\bibinfo  {journal} {Phys. Rev.}\ }\textbf {\bibinfo
  {volume} {D83}},\ \bibinfo {pages} {124005} (\bibinfo {year} {2011})},\
  \Eprint {http://arxiv.org/abs/1104.4790} {arXiv:1104.4790 [gr-qc]}
  \BibitemShut {NoStop}%
\bibitem [{\citenamefont {Ayon-Beato}\ and\ \citenamefont
  {Garcia}(1998)}]{AyonBeato:1998ub}%
  \BibitemOpen
  \bibfield  {author} {\bibinfo {author} {\bibfnamefont {E.}~\bibnamefont
  {Ayon-Beato}}\ and\ \bibinfo {author} {\bibfnamefont {A.}~\bibnamefont
  {Garcia}},\ }\href {\doibase 10.1103/PhysRevLett.80.5056} {\bibfield
  {journal} {\bibinfo  {journal} {Phys. Rev. Lett.}\ }\textbf {\bibinfo
  {volume} {80}},\ \bibinfo {pages} {5056} (\bibinfo {year} {1998})},\ \Eprint
  {http://arxiv.org/abs/gr-qc/9911046} {arXiv:gr-qc/9911046 [gr-qc]}
  \BibitemShut {NoStop}%
\bibitem [{\citenamefont {Ayón-Beato}\ and\ \citenamefont
  {Garcı́a}(1999)}]{AyónBeato199925}%
  \BibitemOpen
  \bibfield  {author} {\bibinfo {author} {\bibfnamefont {E.}~\bibnamefont
  {Ayón-Beato}}\ and\ \bibinfo {author} {\bibfnamefont {A.}~\bibnamefont
  {Garcı́a}},\ }\href {\doibase
  http://dx.doi.org/10.1016/S0370-2693(99)01038-2} {\bibfield  {journal}
  {\bibinfo  {journal} {Phys. Let. B}\ }\textbf {\bibinfo {volume} {464}},\
  \bibinfo {pages} {25 } (\bibinfo {year} {1999})}\BibitemShut {NoStop}%
\bibitem [{\citenamefont {Ayon-Beato}\ and\ \citenamefont
  {Garcia}(1999)}]{AyonBeato:1999ec}%
  \BibitemOpen
  \bibfield  {author} {\bibinfo {author} {\bibfnamefont {E.}~\bibnamefont
  {Ayon-Beato}}\ and\ \bibinfo {author} {\bibfnamefont {A.}~\bibnamefont
  {Garcia}},\ }\href {\doibase 10.1023/A:1026640911319} {\bibfield  {journal}
  {\bibinfo  {journal} {Gen. Rel. Grav.}\ }\textbf {\bibinfo {volume} {31}},\
  \bibinfo {pages} {629} (\bibinfo {year} {1999})},\ \Eprint
  {http://arxiv.org/abs/gr-qc/9911084} {arXiv:gr-qc/9911084 [gr-qc]}
  \BibitemShut {NoStop}%
\bibitem [{Note1()}]{Note1}%
  \BibitemOpen
  \bibinfo {note} {Actually, a general static, spherically symmetric solution
  with an electric charge in GR coupled to NLED has been obtained long before
  \cite {Pellicer1969}}\BibitemShut {NoStop}%
\bibitem [{\citenamefont {Bronnikov}\ \emph {et~al.}(1979)\citenamefont
  {Bronnikov}, \citenamefont {Melnikov}, \citenamefont {Shikin},\ and\
  \citenamefont {Staniukovich}}]{BRONNIKOV197984}%
  \BibitemOpen
  \bibfield  {author} {\bibinfo {author} {\bibfnamefont {K.}~\bibnamefont
  {Bronnikov}}, \bibinfo {author} {\bibfnamefont {V.}~\bibnamefont {Melnikov}},
  \bibinfo {author} {\bibfnamefont {G.}~\bibnamefont {Shikin}}, \ and\ \bibinfo
  {author} {\bibfnamefont {K.}~\bibnamefont {Staniukovich}},\ }\href {\doibase
  http://dx.doi.org/10.1016/0003-4916(79)90235-5} {\bibfield  {journal}
  {\bibinfo  {journal} {Annals of Physics}\ }\textbf {\bibinfo {volume}
  {118}},\ \bibinfo {pages} {84 } (\bibinfo {year} {1979})}\BibitemShut
  {NoStop}%
\bibitem [{\citenamefont {Bronnikov}(2000)}]{Bronnikov2000}%
  \BibitemOpen
  \bibfield  {author} {\bibinfo {author} {\bibfnamefont {K.~A.}\ \bibnamefont
  {Bronnikov}},\ }\href {\doibase 10.1103/PhysRevLett.85.4641} {\bibfield
  {journal} {\bibinfo  {journal} {Phys. Rev. Lett.}\ }\textbf {\bibinfo
  {volume} {85}},\ \bibinfo {pages} {4641} (\bibinfo {year}
  {2000})}\BibitemShut {NoStop}%
\bibitem [{\citenamefont {Bronnikov}(2001)}]{Bronnikov2001}%
  \BibitemOpen
  \bibfield  {author} {\bibinfo {author} {\bibfnamefont {K.~A.}\ \bibnamefont
  {Bronnikov}},\ }\href {\doibase 10.1103/PhysRevD.63.044005} {\bibfield
  {journal} {\bibinfo  {journal} {Phys. Rev. D}\ }\textbf {\bibinfo {volume}
  {63}},\ \bibinfo {pages} {044005} (\bibinfo {year} {2001})}\BibitemShut
  {NoStop}%
\bibitem [{\citenamefont {Flachi}\ and\ \citenamefont
  {Lemos}(2013)}]{Flachi:2012nv}%
  \BibitemOpen
  \bibfield  {author} {\bibinfo {author} {\bibfnamefont {A.}~\bibnamefont
  {Flachi}}\ and\ \bibinfo {author} {\bibfnamefont {J.~P.~S.}\ \bibnamefont
  {Lemos}},\ }\href {\doibase 10.1103/PhysRevD.87.024034} {\bibfield  {journal}
  {\bibinfo  {journal} {Phys. Rev.}\ }\textbf {\bibinfo {volume} {D87}},\
  \bibinfo {pages} {024034} (\bibinfo {year} {2013})},\ \Eprint
  {http://arxiv.org/abs/1211.6212} {arXiv:1211.6212 [gr-qc]} \BibitemShut
  {NoStop}%
\bibitem [{\citenamefont {Li}\ \emph {et~al.}(2013)\citenamefont {Li},
  \citenamefont {Ma},\ and\ \citenamefont {Lin}}]{Li:2013fka}%
  \BibitemOpen
  \bibfield  {author} {\bibinfo {author} {\bibfnamefont {J.}~\bibnamefont
  {Li}}, \bibinfo {author} {\bibfnamefont {H.}~\bibnamefont {Ma}}, \ and\
  \bibinfo {author} {\bibfnamefont {K.}~\bibnamefont {Lin}},\ }\href {\doibase
  10.1103/PhysRevD.88.064001} {\bibfield  {journal} {\bibinfo  {journal} {Phys.
  Rev.}\ }\textbf {\bibinfo {volume} {D88}},\ \bibinfo {pages} {064001}
  (\bibinfo {year} {2013})},\ \Eprint {http://arxiv.org/abs/1308.6499}
  {arXiv:1308.6499 [gr-qc]} \BibitemShut {NoStop}%
\bibitem [{\citenamefont {Toshmatov}\ \emph {et~al.}(2015)\citenamefont
  {Toshmatov}, \citenamefont {Abdujabbarov}, \citenamefont {Stuchlik},\ and\
  \citenamefont {Ahmedov}}]{Toshmatov:2015wga}%
  \BibitemOpen
  \bibfield  {author} {\bibinfo {author} {\bibfnamefont {B.}~\bibnamefont
  {Toshmatov}}, \bibinfo {author} {\bibfnamefont {A.}~\bibnamefont
  {Abdujabbarov}}, \bibinfo {author} {\bibfnamefont {Z.}~\bibnamefont
  {Stuchlik}}, \ and\ \bibinfo {author} {\bibfnamefont {B.}~\bibnamefont
  {Ahmedov}},\ }\href {\doibase 10.1103/PhysRevD.91.083008} {\bibfield
  {journal} {\bibinfo  {journal} {Phys. Rev.}\ }\textbf {\bibinfo {volume}
  {D91}},\ \bibinfo {pages} {083008} (\bibinfo {year} {2015})},\ \Eprint
  {http://arxiv.org/abs/1503.05737} {arXiv:1503.05737 [gr-qc]} \BibitemShut
  {NoStop}%
\bibitem [{\citenamefont {Fernando}\ and\ \citenamefont
  {Correa}(2012)}]{PhysRevD.86.064039}%
  \BibitemOpen
  \bibfield  {author} {\bibinfo {author} {\bibfnamefont {S.}~\bibnamefont
  {Fernando}}\ and\ \bibinfo {author} {\bibfnamefont {J.}~\bibnamefont
  {Correa}},\ }\href {\doibase 10.1103/PhysRevD.86.064039} {\bibfield
  {journal} {\bibinfo  {journal} {Phys. Rev. D}\ }\textbf {\bibinfo {volume}
  {86}},\ \bibinfo {pages} {064039} (\bibinfo {year} {2012})}\BibitemShut
  {NoStop}%
\bibitem [{\citenamefont {Xi}\ and\ \citenamefont {Ao}(2016)}]{Xi:2016qrg}%
  \BibitemOpen
  \bibfield  {author} {\bibinfo {author} {\bibfnamefont {P.}~\bibnamefont
  {Xi}}\ and\ \bibinfo {author} {\bibfnamefont {X.-c.}\ \bibnamefont {Ao}},\
  }\href {\doibase 10.1007/s10714-016-2017-6} {\bibfield  {journal} {\bibinfo
  {journal} {Gen. Rel. Grav.}\ }\textbf {\bibinfo {volume} {48}},\ \bibinfo
  {pages} {14} (\bibinfo {year} {2016})}\BibitemShut {NoStop}%
\bibitem [{\citenamefont {Bekenstein}(1973)}]{Bekenstein1973a}%
  \BibitemOpen
  \bibfield  {author} {\bibinfo {author} {\bibfnamefont {J.~D.}\ \bibnamefont
  {Bekenstein}},\ }\href {\doibase 10.1103/PhysRevD.7.949} {\bibfield
  {journal} {\bibinfo  {journal} {Phys. Rev. D}\ }\textbf {\bibinfo {volume}
  {7}},\ \bibinfo {pages} {949} (\bibinfo {year} {1973})}\BibitemShut {NoStop}%
\bibitem [{\citenamefont {Hod}(2013)}]{Hod20131489}%
  \BibitemOpen
  \bibfield  {author} {\bibinfo {author} {\bibfnamefont {S.}~\bibnamefont
  {Hod}},\ }\href {\doibase http://dx.doi.org/10.1016/j.physletb.2012.12.013}
  {\bibfield  {journal} {\bibinfo  {journal} {Phys. Lett. B}\ }\textbf
  {\bibinfo {volume} {718}},\ \bibinfo {pages} {1489 } (\bibinfo {year}
  {2013})}\BibitemShut {NoStop}%
\bibitem [{\citenamefont {Hod}(2015)}]{Hod2015prd}%
  \BibitemOpen
  \bibfield  {author} {\bibinfo {author} {\bibfnamefont {S.}~\bibnamefont
  {Hod}},\ }\href {\doibase 10.1103/PhysRevD.91.044047} {\bibfield  {journal}
  {\bibinfo  {journal} {Phys. Rev. D}\ }\textbf {\bibinfo {volume} {91}},\
  \bibinfo {pages} {044047} (\bibinfo {year} {2015})}\BibitemShut {NoStop}%
\bibitem [{\citenamefont {Li}(2013)}]{Li:2013jna}%
  \BibitemOpen
  \bibfield  {author} {\bibinfo {author} {\bibfnamefont {R.}~\bibnamefont
  {Li}},\ }\href {\doibase 10.1103/PhysRevD.88.127901} {\bibfield  {journal}
  {\bibinfo  {journal} {Phys. Rev.}\ }\textbf {\bibinfo {volume} {D88}},\
  \bibinfo {pages} {127901} (\bibinfo {year} {2013})},\ \Eprint
  {http://arxiv.org/abs/1310.3587} {arXiv:1310.3587 [gr-qc]} \BibitemShut
  {NoStop}%
\bibitem [{\citenamefont {Pani}\ \emph {et~al.}(2012)\citenamefont {Pani},
  \citenamefont {Cardoso}, \citenamefont {Gualtieri}, \citenamefont {Berti},\
  and\ \citenamefont {Ishibashi}}]{PhysRevD.86.104017}%
  \BibitemOpen
  \bibfield  {author} {\bibinfo {author} {\bibfnamefont {P.}~\bibnamefont
  {Pani}}, \bibinfo {author} {\bibfnamefont {V.}~\bibnamefont {Cardoso}},
  \bibinfo {author} {\bibfnamefont {L.}~\bibnamefont {Gualtieri}}, \bibinfo
  {author} {\bibfnamefont {E.}~\bibnamefont {Berti}}, \ and\ \bibinfo {author}
  {\bibfnamefont {A.}~\bibnamefont {Ishibashi}},\ }\href {\doibase
  10.1103/PhysRevD.86.104017} {\bibfield  {journal} {\bibinfo  {journal} {Phys.
  Rev. D}\ }\textbf {\bibinfo {volume} {86}},\ \bibinfo {pages} {104017}
  (\bibinfo {year} {2012})}\BibitemShut {NoStop}%
\bibitem [{\citenamefont {Dolan}(2013)}]{2013-Dolan-p124026-124026}%
  \BibitemOpen
  \bibfield  {author} {\bibinfo {author} {\bibfnamefont {S.~R.}\ \bibnamefont
  {Dolan}},\ }\href {\doibase 10.1103/PhysRevD.87.124026} {\bibfield  {journal}
  {\bibinfo  {journal} {Phys. Rev. D}\ }\textbf {\bibinfo {volume} {87}},\
  \bibinfo {pages} {124026} (\bibinfo {year} {2013})}\BibitemShut {NoStop}%
\bibitem [{\citenamefont {Pani}(2013)}]{Pani:2013pma}%
  \BibitemOpen
  \bibfield  {author} {\bibinfo {author} {\bibfnamefont {P.}~\bibnamefont
  {Pani}},\ }\bibfield  {booktitle} {\emph {\bibinfo {booktitle} {{Proceedings,
  Spring School on Numerical Relativity and High Energy Physics (NR/HEP2)}}},\
  }\href {\doibase 10.1142/S0217751X13400186} {\bibfield  {journal} {\bibinfo
  {journal} {Int. J. Mod. Phys.}\ }\textbf {\bibinfo {volume} {A28}},\ \bibinfo
  {pages} {1340018} (\bibinfo {year} {2013})},\ \Eprint
  {http://arxiv.org/abs/1305.6759} {arXiv:1305.6759 [gr-qc]} \BibitemShut
  {NoStop}%
\bibitem [{\citenamefont {Kojima}(1992)}]{Kojima1992}%
  \BibitemOpen
  \bibfield  {author} {\bibinfo {author} {\bibfnamefont {Y.}~\bibnamefont
  {Kojima}},\ }\href {\doibase 10.1103/PhysRevD.46.4289} {\bibfield  {journal}
  {\bibinfo  {journal} {Phys. Rev. D}\ }\textbf {\bibinfo {volume} {46}},\
  \bibinfo {pages} {4289} (\bibinfo {year} {1992})}\BibitemShut {NoStop}%
\bibitem [{\citenamefont {Csizmadia}\ \emph {et~al.}(2013)\citenamefont
  {Csizmadia}, \citenamefont {László},\ and\ \citenamefont
  {Rácz}}]{Csizmadia2013On}%
  \BibitemOpen
  \bibfield  {author} {\bibinfo {author} {\bibfnamefont {P.}~\bibnamefont
  {Csizmadia}}, \bibinfo {author} {\bibfnamefont {A.}~\bibnamefont {László}},
  \ and\ \bibinfo {author} {\bibfnamefont {I.}~\bibnamefont {Rácz}},\ }\href
  {http://stacks.iop.org/0264-9381/30/i=1/a=015010} {\bibfield  {journal}
  {\bibinfo  {journal} {Classical and Quantum Gravity}\ }\textbf {\bibinfo
  {volume} {30}},\ \bibinfo {pages} {015010} (\bibinfo {year}
  {2013})}\BibitemShut {NoStop}%
\bibitem [{\citenamefont {Gundlach}\ \emph {et~al.}(1994)\citenamefont
  {Gundlach}, \citenamefont {Price},\ and\ \citenamefont
  {Pullin}}]{Gundlach1994prd}%
  \BibitemOpen
  \bibfield  {author} {\bibinfo {author} {\bibfnamefont {C.}~\bibnamefont
  {Gundlach}}, \bibinfo {author} {\bibfnamefont {R.~H.}\ \bibnamefont {Price}},
  \ and\ \bibinfo {author} {\bibfnamefont {J.}~\bibnamefont {Pullin}},\ }\href
  {\doibase 10.1103/PhysRevD.49.883} {\bibfield  {journal} {\bibinfo  {journal}
  {Phys. Rev. D}\ }\textbf {\bibinfo {volume} {49}},\ \bibinfo {pages} {883}
  (\bibinfo {year} {1994})}\BibitemShut {NoStop}%
\bibitem [{\citenamefont {Berti}\ \emph {et~al.}(2007)\citenamefont {Berti},
  \citenamefont {Cardoso}, \citenamefont {Gonzalez},\ and\ \citenamefont
  {Sperhake}}]{berti_mining_2007}%
  \BibitemOpen
  \bibfield  {author} {\bibinfo {author} {\bibfnamefont {E.}~\bibnamefont
  {Berti}}, \bibinfo {author} {\bibfnamefont {V.}~\bibnamefont {Cardoso}},
  \bibinfo {author} {\bibfnamefont {J.~A.}\ \bibnamefont {Gonzalez}}, \ and\
  \bibinfo {author} {\bibfnamefont {U.}~\bibnamefont {Sperhake}},\ }\href
  {\doibase 10.1103/PhysRevD.75.124017} {\bibfield  {journal} {\bibinfo
  {journal} {Phys. Rev. D}\ }\textbf {\bibinfo {volume} {75}},\ \bibinfo
  {pages} {124017} (\bibinfo {year} {2007})},\ \Eprint
  {http://arxiv.org/abs/gr-qc/0701086} {arXiv:gr-qc/0701086 [gr-qc]}
  \BibitemShut {NoStop}%
\bibitem [{\citenamefont {Richartz}\ \emph {et~al.}(2009)\citenamefont
  {Richartz}, \citenamefont {Weinfurtner}, \citenamefont {Penner},\ and\
  \citenamefont {Unruh}}]{Richartz:2009mi}%
  \BibitemOpen
  \bibfield  {author} {\bibinfo {author} {\bibfnamefont {M.}~\bibnamefont
  {Richartz}}, \bibinfo {author} {\bibfnamefont {S.}~\bibnamefont
  {Weinfurtner}}, \bibinfo {author} {\bibfnamefont {A.~J.}\ \bibnamefont
  {Penner}}, \ and\ \bibinfo {author} {\bibfnamefont {W.~G.}\ \bibnamefont
  {Unruh}},\ }\href {\doibase 10.1103/PhysRevD.80.124016} {\bibfield  {journal}
  {\bibinfo  {journal} {Phys. Rev.}\ }\textbf {\bibinfo {volume} {D80}},\
  \bibinfo {pages} {124016} (\bibinfo {year} {2009})},\ \Eprint
  {http://arxiv.org/abs/0909.2317} {arXiv:0909.2317 [gr-qc]} \BibitemShut
  {NoStop}%
\bibitem [{\citenamefont {Pellicer}\ and\ \citenamefont
  {Torrence}(1969)}]{Pellicer1969}%
  \BibitemOpen
  \bibfield  {author} {\bibinfo {author} {\bibfnamefont {R.}~\bibnamefont
  {Pellicer}}\ and\ \bibinfo {author} {\bibfnamefont {R.~J.}\ \bibnamefont
  {Torrence}},\ }\href {\doibase http://dx.doi.org/10.1063/1.1665019}
  {\bibfield  {journal} {\bibinfo  {journal} {Journal of Mathematical Physics}\
  }\textbf {\bibinfo {volume} {10}},\ \bibinfo {pages} {1718} (\bibinfo {year}
  {1969})}\BibitemShut {NoStop}%
\end{thebibliography}%
\end{document}